\documentstyle[12pt]{article}
\let\section=\subsection     \let\subsection=\subsubsection                

\begin{document}
\begin{center}
   {\large \bf GLOBAL TRANSMISSION COEFFICIENTS IN
    HAUSER--FESHBACH CALCULATIONS FOR ASTROPHYSICS}\\[5mm]
   T.~RAUSCHER\footnote{APART fellow of the Austrian Academy of
   Sciences}\\[5mm]
   {\small \it  Institut f\"ur Physik, Universit\"at Basel \\
   Klingelbergstr.\ 82, CH--4056 Basel \\[8mm] }
\end{center}

\begin{abstract}\noindent
   The current status of optical potentials employed in
   the prediction of thermonuclear reaction rates
   for astrophysics in the Hauser--Feshbach formalism is discussed.
   Special emphasis is put on $\alpha$+nucleus potentials.
   Further experimental efforts are motivated.
\end{abstract}

\section{Introduction}

The investigation of explosive nuclear burning in astrophysical
environments is a challenge for both theoretical and experimental nuclear
physicists. Highly unstable nuclei are produced in such processes which
again can be targets for subsequent reactions. Cross sections and
astrophysical reaction rates for a large number of nuclei
are required to perform complete network
calculations which take into account all possible reaction links and do
not postulate a priori simplifications. 

The majority of reactions can be described in the framework of the
statistical model (compound nucleus mechanism, Hauser--Feshbach
approach, HF)~\cite{hau52}, provided that the level density of the compound nucleus
is sufficiently large in the contributing energy window~\cite{rau97}.
In astrophysical applications usually different aspects are emphasized
than in pure nuclear physics investigations. Many of
the latter in this long and well established field were focused on
specific reactions, where all or most "ingredients", like optical potentials for
particle transmission coefficients, level densities, resonance
energies and widths of giant resonances to be implemented in
predicting E1 and M1 $\gamma$--transitions, were deduced from experiments.
As long as the statistical model prerequisites are met, this will produce
highly accurate cross sections.
For the majority of nuclei in astrophysical applications such
information is not available. The real challenge is thus not the 
well--established statistical model, but rather to provide all these necessary 
ingredients
in as reliable a way as possible, also for nuclei where none of such
information is available.

\section{Transmission Coefficients}

The final quantities entering the expression for the cross
section in the statistical model~\cite{hau52} are the averaged transmission
coefficients. They do not reflect a resonance behavior but rather
describe absorption via an imaginary part in the (optical)
nucleon--nucleus potential~\cite{mah79}. In astrophysics, usually
reactions induced by light projectiles (neutrons, protons, $\alpha$
particles) are
most important. Global optical potentials are quite well defined for
neutrons and protons. It was shown~\cite{thi83,cow91} that the best fit 
of s--wave neutron strength functions is obtained with the optical
potential by~\cite{jeu77}, based on microscopic
infinite nuclear matter calculations for a given density,
applied with a local density approximation.
It includes corrections of the imaginary part~\cite{fantoni81,mahaux82}.
A similar description is used for protons.
Deformed nuclei are treated by an effective spherical potential of equal
volume~\cite{cow91}.
For a detailed description of the formalism used to calculate E1 and M1
$\gamma$--transmission coefficients and how to include width fluctuation
corrections, see~\cite{rau97,cow91} and references therein.

\subsection{$\alpha$+Nucleus Potentials}

Currently, there are only few global parametrizations for
optical $\alpha$+nucleus potentials at astrophysical energies. 
Most global potentials are of the Saxon--Woods form,
parametrized at energies above about 70 MeV,
e.g.~\cite{sin76,nol87}.
The high Coloumb barrier makes a
direct experimental approach very difficult at low energies. More
recently, there were attempts to extend those parametrizations to
energies below 70 MeV~\cite{avr94}.

Early astrophysical statistical model calculations~\cite{arn72,woo78} made use
of simplified equivalent square well potentials and the black nucleus
approximation. Improved
calculations~\cite{thi87} employed a phenomenological Saxon--Woods
potential~\cite{mann78}, based on extensive data~\cite{mcf66}.
This potential is an energy-- and mass--independent mean
potential. However, especially at low energies the imaginary part of the
potential should be highly energy--dependent. 

Most recent experimental
investigations~\cite{mohr94,atz96} found a systematic mass-- and
energy--dependence and were very successful in describing experimental
scattering data, as well as bound and quasi--bound states
and $B(\rm{E2})$ values,
with folding potentials~\cite{sat79}.
Motivated by that description (a global pa\-ra\-me\-tri\-za\-tion is not
given in~\cite{atz96}), the following global 
$\alpha$+nucleus potential is proposed. The real part 
$V=\lambda V_{\rm f}$ of the optical potential 
is calculated by a double--folding procedure:
\begin{equation}
V_{\rm f}(r) =
  \int \int \rho_{\rm P}(r_{\rm P}) \, \rho_{\rm T}(r_{\rm T}) \,
  v_{\rm eff}(E,\rho_{\rm N} = \rho_{\rm P} + \rho_{\rm T},
        s = |\vec{r}+\vec{r_{\rm P}}-\vec{r_{\rm T}}|) \,
  d^3r_{\rm P} \, d^3r_{\rm T} \quad,
\label{eq:fold}
\end{equation}
where $\rho_{\rm P}$, $\rho_{\rm T}$ are the nuclear densities of projectile and
target, respectively,
and $v_{\rm eff}$ is the effective nucleon--nucleon interaction taken in
the well--established DDM3Y parametrization~\cite{kobos84}. The remaining
strength parameter $\lambda$ can be determined from the 
systematics~\cite{mohr94} of
the volume integral per interacting particle pair
\begin{equation}
\label{vi}
J_R = \frac{1}{A_{\rm P} A_{\rm T}} \, \int V(r) \, d^3r \quad.
\end{equation}
Introducing a mass--dependence in addition to the
energy--dependence, one can reproduce the behavior derived from the
imaginary part by applying the dispersion relation~\cite{atz96}, by
the relation
\begin{equation}
\label{volint}
J_R = \left\{ \begin{array}{ll}
       f(A) + 0.67 E_{\rm c.m.} 
                & (E_{\rm{c.m.}} \le 26) \\
         \left[0.011f(A)-2.847\right]E_{\rm{c.m.}}-1.277f(A)+626.591
                & (26 < E_{\rm{c.m.}} < 120)\,, \\
        \end{array} \right.
\end{equation}
with $f(A)=311.0132\exp\left(A_{\rm T}^{-2/3}\right)$ and
$E_{\rm c.m.}$ given in MeV. Using Eqs.~\ref{vi} and \ref{volint}, 
the value of $\lambda$ can be set.

The mass-- and energy--dependence of the imaginary part is more crucial
for the calculation of transmission coefficients. 
The volume integral of the imaginary part can be parametrized according
to~\cite{bro81}:
\begin{equation}
J_I(E_{\rm{c.m.}}) = \left\{ \begin{array}{rll}
        & \multicolumn{1}{c}{0}
                & {\mbox{for~}} E_{\rm{c.m.}} \le E_0 \\
        & J_0 \cdot \frac{(E_{\rm{c.m.}} - E_0)^2}
                {(E_{\rm{c.m.}} - E_0)^2 + \Delta^2}
                & {\mbox{for~}} E_{\rm{c.m.}} > E_0 \quad,\\
        \end{array} \right.
\label{eq:br}
\end{equation}
where $E_0$ is the threshold energy for inelastic channels,
$J_0$ is a saturation parameter and $\Delta$ the
rise parameter.

As can be seen~\cite{atz96}, the resulting volume integrals are quite
diverse for different nuclei and a straight--forward mass--dependent
parametrization is not possible.
In Ref.~\cite{atz96}, $J_0$ and $\Delta$ were fitted to experimental
data. For a global parametrization, one has to include nuclear structure
and deformation information into the fit.
Having chosen the (probably energy--dependent)
geometry parameters~\cite{avr94} (and
appropriately modified them for deformed nuclei), the
depth $W$ of an imaginary Saxon--Woods potential can be determined with
Eq.~\ref{eq:br}. At energies $E_{\rm c.m.} > 100$ MeV, the value of $W$
should reach the saturation value~\cite{nol87}. This determines the
parameter $J_0$. The threshold $E_0=E_0(\rho)$ is given by 
the energy of the first
excited state. The increase in the number of inelastic (competing)
channels is described by $\Delta$. As a first estimate, this can be
related to the increase in the total number $n=n(\rho)$ of levels:
$ \Delta = f\left(dn/dE\right)= \alpha E_0 + \beta A^{2/3} + \gamma
\ln \left(dn/dE\right)$.
The functional dependence is found by comparison with experimental
values for $\Delta$~\cite{atz96}. Thus, microscopic and deformation effects are
implicitly included via the level density $\rho=\rho(E)$.

With nuclear level densities $\rho(E)$ taken from~\cite{rau97},
a good fit is obtained to the values given in~\cite{atz96}, as well as
for various ($\alpha$,$\gamma$) and (n,$\alpha$) data~\cite{hansper}. 
However, it has to be
emphasized that the aim is to obtain a reasonable {\it global} potential,
i.e.\ with a low {\it average} deviation from experiment. For specific
reactions, such a potential might be improved by fine--tuning to a particular
nucleus (e.g.~\cite{mohr97,som98}) but will lose predictive power by that.

\section{Conclusion}

With the recent improvements of the level density treatment~\cite{rau97}
and the $\alpha$+nucleus potential, the two major remaining uncertainties in
the existing global astrophysical reaction rate calculations have been
attacked. The new descriptions will provide more reliable predictions of
the low--energy cross sections of unstable nuclei.

Nevertheless, to check and further improve current parametrizations, not
only of $\alpha$+nucleus potentials but of all involved quantities,
experimental
data is needed. Especially investigations over a large mass range would
prove useful to fill in gaps in the knowledge of the nuclear structure
of many isotopes and to construct more powerful parameter systematics.
Such investigations should include neutron--
and proton--strength functions, as well as radiative widths, and
charged particle scattering and reaction cross sections for {\it stable}
and unstable isotopes.
This information can be used to make future
large--scale statistical model calculations even more accurate.


\begin{thebibliography}{99}
\itemsep=0cm
\bibitem{hau52}
W. Hauser and H. Feshbach, Phys.\ Rev.\ {\bf 87} (1952) 366
\bibitem{rau97}
T. Rauscher, F.-K. Thielemann, and K.-L. Kratz, Phys.\ Rev.\ C {\bf 56}
(1997) 1613
\bibitem{mah79}
C. Mahaux, H.A. Weidenm\"uller, Ann.\ Rev.\ Part.\ Nucl.\ Sci.\ {\bf
29} (1979) 1
\bibitem{thi83}
F.-K. Thielemann, J. Metzinger, and H.V. Klapdor, Z. Phys.\ A {\bf 309}
(1983) 301
\bibitem{cow91}
J.J. Cowan, F.-K. Thielemann, J.W. Truran, Phys.\ Rep.\ {\bf 208}
(1991) 267
\bibitem{jeu77}
J.P. Jeukenne, A. Lejeune and C. Mahaux, Phys.\ Rev.\ C {\bf 16} (1977)
80
\bibitem{fantoni81}
S. Fantoni, B.L. Friman and V.R. Pandharipande, Phys.\ Rev.\ Lett.\ {\bf
48} (1981) 1089
\bibitem{mahaux82}
C. Mahaux, Phys.\ Rev.\ C {\bf 82} (1982) 1848
\bibitem{sin76}
P.P. Singh and P. Schwandt, Nukleonika {\bf 21} (1976) 451
\bibitem{nol87}
M. Nolte, H. Machner and J. Bojowald, Phys.\ Rev.\ C {\bf 36} (1987)
1312
\bibitem{avr94}
V. Avrigeanu, P.E. Hodgson, M. Avrigeanu, Phys.\ Rev.\ C {\bf 49}
(1994) 2136
\bibitem{arn72}
M. Arnould, A\&A {\bf 19} (1972) 92
\bibitem{woo78}
S.E. Woosley, W.A. Fowler, J.A. Holmes, and B.A. Zimmerman, At.\ Data
Nucl.\ Data Tables {\bf 22} (1978) 371
\bibitem{thi87}
F.-K. Thielemann, M. Arnould and J.W. Truran, in {\it Advances in
Nuclear Astrophysics}, ed.\ E. Vangioni--Flam, Gif sur Yvette 1987, Editions
Fronti\`ere, p.\ 525
\bibitem{mann78}
F.M. Mann, 1978, Hanford Engineering, report HEDL-TME 78--83
\bibitem{mcf66}
L. McFadden and G.R. Satchler, Nucl.\ Phys.\ {\bf 84} (1966) 177
\bibitem{mohr94}
P. Mohr, H. Abele, U. Atzrott, G. Staudt,
        R. Bieber, K. Gr\"un, H. Oberhummer, T. Rauscher, and
E. Somorjai, in
        {\it Proc.\ Europ.\ Workshop on Heavy Element Nucleosynthesis},
       eds.\ E. Somorjai and Zs.\ F\"ul\"op, Institute of Nuclear
Research, Debrecen 1994, p.\ 176
\bibitem{atz96}
U. Atzrott, P. Mohr, H. Abele, C. Hillenmayer, and
        G. Staudt,
        Phys.\ Rev.\ C {\bf 53} (1996) 1336
\bibitem{sat79}
G.R. Satchler and W.G. Love,
        Phys.~Rep.~{\bf 55} (1979) 183
\bibitem{kobos84}
A.M. Kobos, B.A. Brown, R. Lindsay, and
        R. Satchler,
        Nucl.\ Phys.\ {\bf A425} (1984) 205
\bibitem{bro81}
G.E. Brown and M. Rho, Nucl.\ Phys.\ {\bf A372} (1981) 397
\bibitem{hansper}
V. Hansper, private communication
\bibitem{mohr97}
P. Mohr, T. Rauscher, H. Oberhummer, Z. M\'at\'e, Zs. F\"ul\"op, E.
Somorjai, M. Jaeger, and G. Staudt, Phys.\ Rev.\ C {\bf 55} (1997) 1523
\bibitem{som98}
E. Somorjai, Zs. F\"ul\"op, \'A.Z. Kiss, C.E. Rolfs, H.P. Trautvetter,
U. Greife, M. Junker, S. Goriely, M. Arnould, M. Rayet, T. Rauscher, and
H. Oberhummer, A\&A, in press
\end{thebibliography}
\end{document}